\documentclass[conference,peer-reviewed]{aesconf}

\graphicspath{{./}{figures/}}

\usepackage[utf8]{inputenc}

\usepackage{microtype}

\usepackage[numbers,square]{natbib}

\usepackage{booktabs}
\usepackage{color}
\usepackage{url}
\usepackage{amsmath}
\usepackage{subcaption}
\usepackage{layouts}
\usepackage{amssymb}

\renewcommand{\exp}[1]{e^{#1}}
\newcommand{\fadein}{\texttt{FADE-IN}}
\newcommand{\posonly}{\texttt{POS-ONLY}}

\title{Fade-in Reverberation in Multi-room Environments\\Using the Common-Slope Model}

\author[1]{Kyung Yun Lee}
\author[1]{Nils Meyer-Kahlen}
\author[1]{Georg Götz}
\author[2]{U. Peter Svensson}
\author[3]{Sebastian J. Schlecht}
\author[1]{Vesa Välimäki}
\affil[1]{Acoustics Lab, Dpt. of Information and Communications Engineering, Aalto University, Finland}
\affil[2]{Dpt. of Electronic Systems, Norwegian University of Science and Technology, Norway}
\affil[3]{Friedrich-Alexander-Universit\"{a}t Erlangen-N\"{u}rnberg (FAU), Germany}

\correspondence{Kyung Yun Lee}{kyung.y.lee@aalto.fi}

\lastnames{Lee et al.}

\shorttitle{Fade-in reverberation}

\savebox{\AEStop}{%
	\begin{minipage}{\textwidth}%
		\rule{\textwidth}{1.5pt}\\%
		\\%
		\begin{minipage}[c][\iftoggle{convention}{3.2cm}{3.7cm}][t]{0\textwidth}%
			\includegraphics[width=20mm]{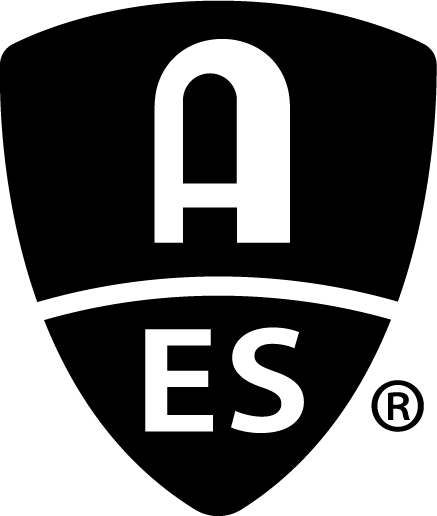}%
		\end{minipage}%
		\begin{minipage}{\textwidth}%
			\sffamily%
			\begin{center}%
				\LARGE Audio Engineering Society\\%
				\iftoggle{express_paper}{%
				\hspace{3mm}\fontsize{36}{38pt}\selectfont Convention Express\\Paper \AESExpressPaperNumber\\%
				}{%
				\iftoggle{convention}{%
				\fontsize{36}{38pt}\selectfont Convention Paper\\%
				}{%
				\fontsize{36}{38pt}\selectfont Conference Paper\\%
				}}%
				\vspace{0.2cm}%
				\large Presented at the AES \ifx\AESConferenceNumber\empty\else\AESConferenceNumber \fi\iftoggle{convention}{Convention\\}{\AESConferencePrefix Conference on\\}%
				\iftoggle{convention}{}{\AESConferenceTopic\\}%
				\AESConferenceDate\ifx\AESConferenceLocation\empty\else, \AESConferenceLocation\fi%
			\end{center}%
		\end{minipage}\\%
		\vspace{0.2cm}\\%
		\begin{minipage}{\textwidth}%
			\rmfamily\itshape\small	\AESLegalTextPrefix\ \AESLegalText%
		\end{minipage}\\%
		\\%
		\rule{\textwidth}{1.5pt}%
	\end{minipage}%
}

\begin{document}

\twocolumn[
\maketitle 

\begin{onecolabstract}
In multi-room environments, modelling the sound propagation is complex due to the coupling of rooms and diverse source-receiver positions. A common scenario is when the source and the receiver are in different rooms without a clear line of sight. For such source-receiver configurations, an initial increase in energy is observed, referred to as the ``fade-in'' of reverberation. Based on recent work of representing inhomogeneous and anisotropic reverberation with common decay times, this work proposes an extended parametric model that enables the modelling of the fade-in phenomenon. The method performs fitting on the envelopes, instead of energy decay functions, and allows negative amplitudes of decaying exponentials. We evaluate the method on simulated and measured multi-room environments, where we show that the proposed approach can now model the fade-ins that were unrealisable with the previous method. 

\end{onecolabstract}
]

\section{Introduction}
Moving throughout our everyday life, we often encounter multi-room environments, where several spaces are coupled to each other. Naturally, it is desirable to model sound propagation through such coupled rooms in augmented and virtual reality (AR/VR) applications. 
When two or more rooms are connected, there is an exchange of sound energy between the rooms through the apertures that connect them \cite{CremerMuller2016PrinciplesAndApplicationsOfRoomAcoustics,Kuttruff2009RA,SuGul2019DiffusionEquationModelingSoundEnergyFlowCoupled}. Depending on the position of the sound source, a room can be directly excited by the sound source or indirectly by the sound originating from the neighbouring rooms. 



\begin{figure*}%
    \centering
    \begin{subfigure}[b]{0.35\textwidth}
        \centering
        \includegraphics[height=6.12cm]{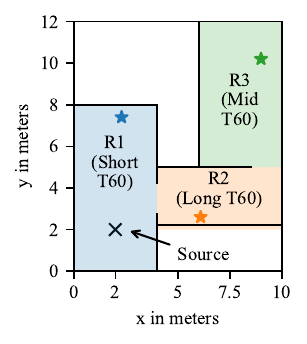}
        \caption{Floor plan}
        \label{fig:simulation_map}
    \end{subfigure}%
    \hfill
    \begin{subfigure}[b]{0.18\textwidth}
        \centering
        \includegraphics[height=6cm]{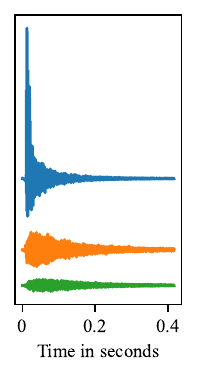}
        \caption{RIRs}
        \label{fig:simulation_rir}
    \end{subfigure}%
    \hfill
    \begin{subfigure}[b]{0.46\textwidth}
        \centering
        \includegraphics[height=6.1cm]{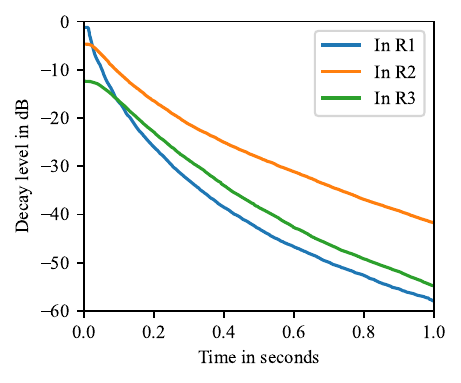}
        \caption{Energy decay curves}
        \label{fig:simulation_edf}
    \end{subfigure}
    \caption{A coupled room example with three rooms from the simulation dataset. The ``$\times$'' indicates the source position and the three stars, $\bigstar$, show the receiver positions.}
    \label{fig:simulation_edf_rir}
\end{figure*}

An example of a multi-room system with three rooms is shown in Fig.~\ref{fig:simulation_edf_rir}. Here, the source is located in the first room (R1), and a receiver is located in each room. 
Inspecting the room impulse responses (RIRs) in Fig.~\ref{fig:simulation_rir}, it becomes clear that only the receiver in R1 exhibits direct sound and monotonically decaying reverberation, which is commonly assumed when synthesizing the acoustics of single rooms. The responses of receivers in R2 and R3 have no direct sound and show an initial increase in the energy. We refer to these responses as room-to-room responses and to the observed energy ramps as the ``fade-in'' of reverberation. In the energy decay curves (EDCs) of RIRs in R2 and R3 (Fig.~\ref{fig:simulation_edf}), this fade-in manifests itself as a convex shape in the initial part, as opposed to a concave one for that in R1. 
The occurrence of such EDC shapes in coupled rooms was already discussed in \cite{Kuttruff2009RA,pu2011different}. 




A common approach for modelling coupled rooms is using the convolution of impulse responses of each room \cite{svensson1998energy, dirk2007hybrid, stavrakis2008topological}. The underlying assumption is that a sound source first excites one room, which, in turn, excites the next room, depending on the size of the aperture between them. This also assumes that the sound returning from the second room is negligible. For several coupled rooms, the propagation can then be seen as a graph, in which the final transfer function is obtained through convolving a series of filters along the trajectory \cite{stavrakis2008topological, wefers2009real, schroder2011raven}.

In addition, various signal-processing structures have been studied to produce non-exponential artificial reverberation, such as multi-slope responses \cite{Piirila1998, Lee2010}, a smooth fade-in of RIRs with a feedback delay network (FDN) \cite{Meyer2020} and FDN designs for coupled rooms \cite{das2021grouped, kirsch2023computationally}. In particular, FDN-based approaches for modelling coupled rooms have been explored due to their efficiency, which is necessary in AR/VR applications. 


Lastly, a very general approach for modelling complex acoustical scenarios is to decompose the energy decay function (EDF), referred to as EDC in a graphical context, of a RIR into a weighted sum of multiple exponential functions with different decay constants, e.g., \cite{Kuttruff2009RA, xiang2001evaluation, summers2004statistical, pu2011different}

\begin{align} 
\text{EDF}(t)=  N \psi_{0}(t) + \sum_{k=1}^\kappa A_{k} \psi_{k}(t) \\ 
\psi_{k}(t) = \begin{cases} 
 L - t, & \text{if $k = 0$} \\
 \text{exp}(\frac{\text{ln}(10^{-6})\, t }{f_\mathrm{s} T_{k}}) , & \text{if $k > 0$}
 \end{cases}
\label{eq:edf2}
\end{align}
where $\kappa$ is the number of decay components, $N$ corresponds to a constant noise term, and $t$ denotes time in samples. For decay kernels $\psi$, $T_k$ are the decay times, $L$ is the length of the RIR in samples, and $f_\mathrm{s}$ refers to the sampling rate. It has been shown that some of the exponential amplitudes $A_k$ are negative when the source and the receiver are in different rooms \cite{Kuttruff2009RA, pu2011different}.
In recent work, Götz et al.~\cite{Goetz2023Common} demonstrated that decay rates can be shared between responses from different positions (and look directions). Then, a complete set of measurements can be modelled by only fitting different amplitudes $A_k$ for each position (and orientation) with a constraint that $A_k$ has to be positive. Once such a model is fit to a set of measurements, it enables efficient reverberator architectures with only as many reverberators as different slopes are found \cite{Goetz2024DynamicAES}.

While this is a powerful model that has been shown to fit many responses, we show here that, in its current form, it cannot model fade-in reverberations of coupled rooms. However, we demonstrate that allowing negative amplitudes and fitting on envelopes instead of on EDFs, overcomes this limitation. The improvement is especially visible for modeling the responses with fade-in, which is most pronounced when there is no line of sight to the source, as for RIRs in R2 and R3 in the introductory example of Fig.~\ref{fig:simulation_edf_rir}.


In the following section, we motivate why the model fit improves when allowing for negative coefficients. Then, in Sec. \ref{sec:method}, we explain how the model fit was performed and in Sec. \ref{sec:evaluation}, we show application of the method to simulated data and a novel dataset of measurements in a large, multi-room space. Sec.~\ref{sec:conclusion} concludes the paper.

\section{Background: Convolution of exponentials}
As discussed above, room-to-room responses can be modelled by convolving the response of one room with the response of the neighbouring room. In statistical acoustics, the response of one room is often represented as a sum of direct sound and exponentially decaying Gaussian noise. If there is no direct sound, the convolution of the two responses can be modelled as 
\begin{align}
    h_\mathrm{R2R}(t) &= h_1(t) *  h_2(t) \\ &= \int_{-\infty}^{\infty} h_1(\tau)   h_2(t-\tau) d\tau \nonumber \\
    &= \int_{-\infty}^{\infty} \text{exp} (-\delta_1 \tau ) n_1(\tau) u(\tau) \nonumber \\ & \quad \quad \quad \text{exp}({-\delta_2 (t-\tau)})n_2(t - \tau) u(t - \tau) d\tau,  \nonumber
\end{align}
where $*$ is the convolution operation, $h_1(t)$ and $h_2(t)$ are the two decaying noise sequences, $n_{1}(t)$ and $n_{2}(t)$ denote realizations of independent Gaussian random variables, $u(t)$ is the Heaviside step function, and $h_\mathrm{R2R}$ is the resulting room-to-room response. As a consequence of this model, $h_\mathrm{R2R}(t)$ is a random variable, with its distribution depending on time $t$.



In \cite{svensson2011properties}, it has been demonstrated that the standard deviation of $h_\mathrm{R2R}(t)$ is equivalent to the difference of exponentials with the same decay rate, i.e., 
\begin{align} 
\sigma{\{ h_\mathrm{R2R}(t) \}}
&= \frac{1}{ \delta_2-\delta_1}  \exp{ -\delta_1 t} - \frac{1}{ \delta_2-\delta_1}  \exp{ -\delta_2 t}, 
\label{eq:r2rdifference}
\end{align}
where $\sigma \{ \cdot \}$ denotes taking the standard deviation. Assuming ergodicity, the standard deviation is, in practice, replaced by a short-term average $s(t)$, computed over a time window with length $l$, equivalent to the envelope.
 \begin{align}
 s(t) = \sqrt{\frac{1}{l}\sum_{i=t-l+1}^t h_\mathrm{R2R}^2(i)}
 \end{align}

The fundamental mathematical result in Eq.~\eqref{eq:r2rdifference} motivates the subsequent development of a model that is able to capture fade-in reverberation. If two coupled rooms have decay rates $\delta_1$ and $\delta_2$, the room-to-room response can be represented with two terms of the model formulation developed in the next section, one with negative and the other with positive amplitude.

\section{Common-slope model with fade-in}
\label{sec:method}

\begin{figure*}[!th]
  \includegraphics[width=\textwidth]{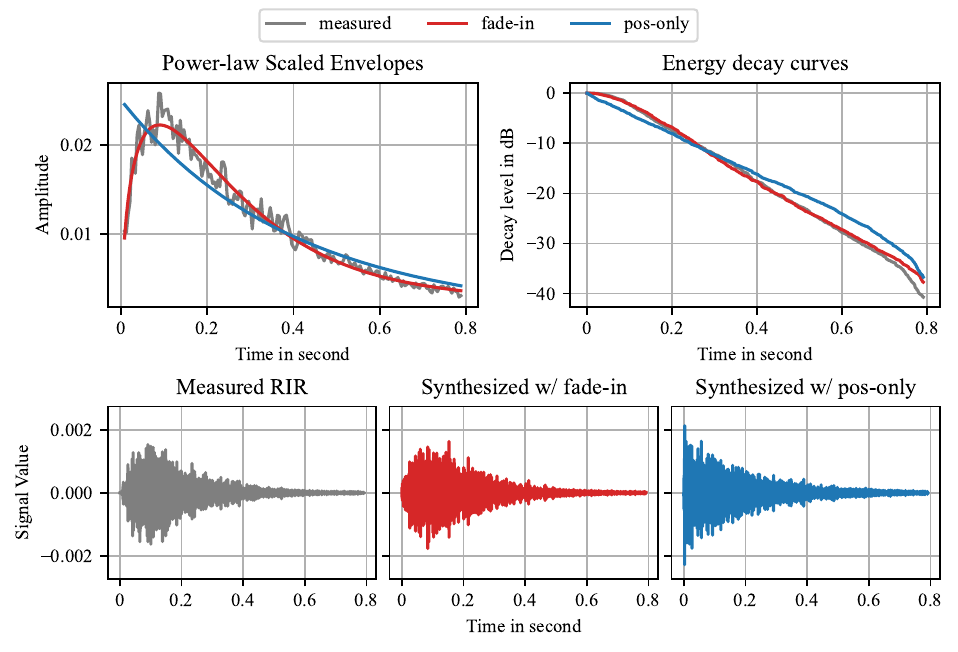}
  \caption{An example of how the \fadein{} and \posonly{} model performs on RIRs with fade-in reverberation. Here, the result of the frequency band centered at 2000 Hz is shown. Note the fitting result on the initial part of RIRs. While the \fadein{} model can capture the fade-in behavior, the \posonly{} model cannot.}
  \label{fig:measuremnt_fit_example1}
\end{figure*} 

The fade-in reverberation model is based on the common-slope model, which describes late reverberation in coupled rooms with a set of common decay times \cite{Goetz2023Common}. Energy envelopes of individual RIRs in a set of measurements can then be represented as a linear combination of decaying exponentials and a noise term. The exponentials decay according to the common decay times, and the amplitude variations across the geometry convey the inhomogeneous and anisotropic characteristics of the late reverberation. In order to obtain the amplitudes of the exponentials, a least-squares fit on the EDFs is performed. 

However, as mentioned above, the major limitation of the common-slope model is its inability to model the fade-in. The limitation comes from the use of EDFs as the fitting domain and the positivity constraint on the amplitudes. When using EDFs, fade-ins are often not pronounced, because the proportion of fade-ins to the rest of the reverberation is very small \cite{summers2004statistical}. In addition, using only positive amplitudes can only model the monotonically decaying behavior, not the initial increase.
Thus, we propose an improved \fadein{} model, which performs fitting on the RIR envelopes, as demonstrated by \cite{karjalainen2002estimation}, and relaxes the optimization constraint to allow negative amplitude values. Despite allowing negative amplitudes, the objective function is still constrained, such that the total sum of the exponentials remains positive.

\begin{figure*}[!th]
  \includegraphics[width=\textwidth]{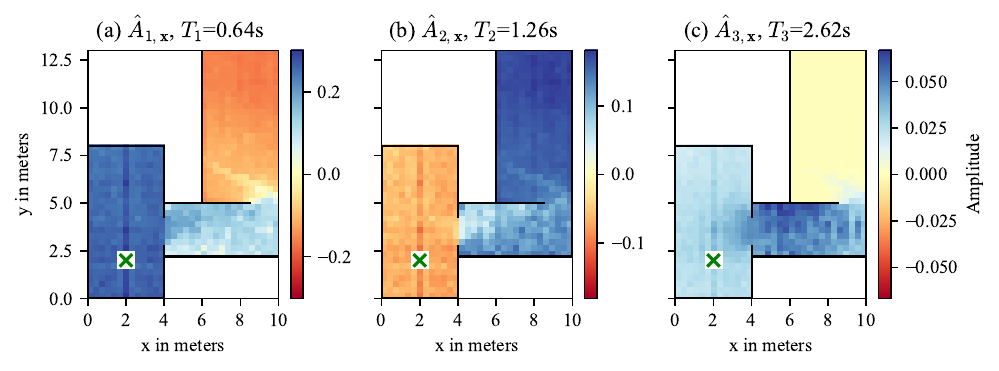}
  \caption{The \fadein{} model result on the simulation dataset, summarizing the results in the 500 Hz, 1 kHz, and 2 kHz octave bands. The source location is marked with the green ``$\times$'' and the receivers are distributed uniformly across the rooms. The amplitudes of each decay time at all receiver positions are shown. The spatial distribution of negative amplitudes can be observed.}
  \label{fig:simulation_result_plot}
\end{figure*}

Akin to the common-slope model formulation \cite{Goetz2023Common}, the \fadein{} model represents the envelope, instead of EDFs, of a RIR as a weighted sum of decaying exponentials. Such representation happens over multiple frequency bands. We choose 7 octave bands from 125 Hz to 8000 Hz. 
The envelope of a RIR at the frequency band, $b$, is formulated as
\begin{align}
s_b(\mathbf{x},t) =  \hat{N}_{b,\mathbf{x}} \psi_{b,0}(t) + \sum_{k=1}^\kappa \hat{A}_{b,k,\mathbf{x}} \psi_{b,k}(t) \\ 
\psi_{b,k}(t) = \begin{cases} 
 1 , & \text{if $k = 0$} \\
 \text{exp}(\frac{\text{ln}(10^{-6})\, t }{f_\mathrm{s} T_{b,k}}) , & \text{if $k > 0$} ,
 \end{cases}
\end{align}
where $\mathbf{x}$ is the source-receiver configuration (locations and directions) 
and $\hat{A}_{b,k,\mathbf{x}}$
and $\hat{N}_{b,\mathbf{x}}$ are the amplitudes of the $k$-th exponential decay envelope and noise at $b$, respectively. 
A set of $T_{b,k}$ are pre-determined by performing K-means clustering over all decay times of octave-filtered RIRs in the same environment \cite{Goetz2023Common}. 
For a specific case of two coupled rooms, room-to-room responses, as described in Eq.~\eqref{eq:r2rdifference}, should lead to pairs of positive and negative coefficients $\hat{A}_{b,k,\mathbf{x}}$ associated with the decay rates of the two rooms. Negative coefficients also appear in environments with more than two coupled rooms, which can be formulated as an extension of Eq.~\eqref{eq:r2rdifference} and are also accomodated in the \fadein{} model with an arbitrary number of decay times $\kappa$. 

In order to obtain the envelope amplitude $\hat{A}_{b,k,\mathbf{x}}$ for a RIR, a nonlinear least squares fit is performed at each $b$
\begin{align}
    \underset{{\hat{A}_{b,k,\mathbf{x}}, \hat{N}_{b,\mathbf{x}}}}{\text{argmin}} \ \sum_{t=0}^{L}[f(\bar{s}_b(\mathbf{x}, t)) - f(s_b(\mathbf{x},t))]^2 \label{eq:objective_function}\\ 
   \text{subject to } s_b(\mathbf{x},t) - \hat{N}_{b,\mathbf{x}}\psi_{b,0}(t) \geq 0    
\end{align}
with a constraint that the estimated envelope values without noise floor must be greater than or equal to zero. 
Here, $\bar{s}_b(\mathbf{x},t)$ is the groundtruth envelope of the original RIR, $h_b(\mathbf{x},t)$. The envelope of a RIR is computed with the given function 
\begin{align}
    \bar{s}_b(\mathbf{x}, t) = g(h_b(\mathbf{x}, t)), 
\end{align}
where $g$ can be any type of downsampling envelope function, obtained by using methods such as Hilbert transform or the moving root-mean-square (RMS). In this work, we use the moving RMS with the non-overlapping window and an additional lowpass filter. At a sampling rate of 48 kHz, the original RIR of 48000 samples is downsampled to have an envelope of length 200 samples. 
The function $f(\cdot)$ applied to the envelopes is a power law scaling function with factor of 0.5, given by $f(y(t)) = \left[ y(t) \right]^{0.5} = \sqrt{y(t)}$. This scaling has been shown to be an effective compromise between the linear and the logarithmic scaling \cite{karjalainen2002estimation}.

With the estimated envelope at a given octave band, the octave-filtered RIR can be synthesized through the multiplication of the envelope with a filtered Gaussian noise, $n_b(t)$.
\begin{align}
    \bar{h}_b(t) = \bar{s}_b(\mathbf{x},t) n_b(t)
\end{align}
Then, the broadband RIR is simply the sum over the results from all bands:  
\begin{align}
    \bar{h}(t) = \sum_{b=1}^B \bar{h}_b(t),
\end{align}
where B is the number of frequency bands.

As a point of comparison, we also fit a model similar to the one discussed in \cite{Goetz2023Common,Goetz2024DynamicAES} and refer to it as \posonly{} model. While using the same objective function as Eq.~(\ref{eq:objective_function}), this model is fitted with a different constraint that the amplitudes of the decaying exponentials must be greater or equal to zero. 
\begin{align}
    \hat{A}_{b,k,\mathbf{x}} >= 0 
\end{align}
Fig.~\ref{fig:measuremnt_fit_example1} shows how the \fadein{} model and the \posonly{} model fit on one of the measured RIRs. It is clear that the fade-in phenomenon can only be captured by allowing negative amplitudes of the decaying exponentials. When restricted to positive amplitudes, the fade-in fitting becomes unattainable, and distorts the estimation of the true EDF. In case of RIRs without fade-in, both models behave similarly.


\begin{figure}[!h]
    \centering
    \subfloat[\centering RMSE of the \fadein{} model]{{\includegraphics[width=\columnwidth]{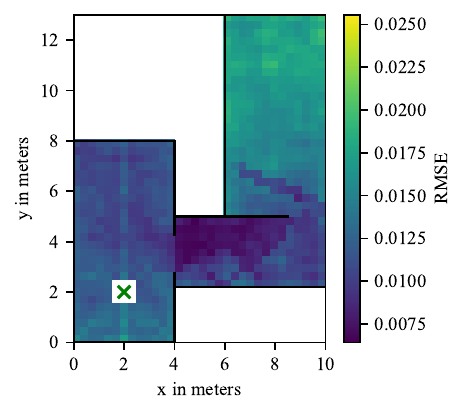} }}%
    \vfill
    \subfloat[\centering RMSE of the \posonly{} model]{{\includegraphics[width=\columnwidth]{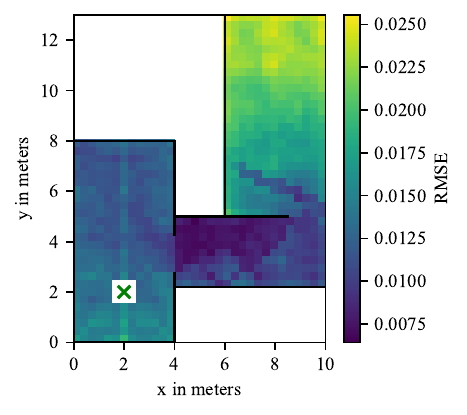} }}%
    \caption{RMSE of model fitted envelopes and the original RIR envelopes at all receiver locations are shown. The source location is marked with ``$\times$''. Plotted values are the sum of RMSE over all 7 octave bands. The top figure shows the result of the \fadein{} model and the bottom shows that of the \posonly{} model. Especially in R3, it can be seen that the \fadein{} model fits better than the \posonly{} model.}%
    \label{fig:simulation_error_plot}
\end{figure}

\section{Evaluation}
\label{sec:evaluation}
We evaluate the proposed \fadein{} model on both simulated and real-world multi-room environment datasets. A comparison with the \posonly{} model is also presented. Listening examples of both the \fadein{} and the \posonly{} model is available at \url{https://kyungyunlee.github.io/fade-in-reverb-demo/}.

\subsection{Simulation dataset}

\subsubsection{Dataset}
The simulated dataset presented in \cite{Goetz2024DynamicAES} consists of RIRs from three coupled rooms as shown in Fig.~\ref{fig:simulation_map}. The total dimension is 10 $\times$ 12 $\times$ 3 m. RIR simulation is done using the Treble suite\footnote{https://www.treble.tech/}. Room 2 is designed as the most reverberant with absorption coefficient $\alpha_{2} = 0.01$, while Rooms 1 and 3 have less reverberation with $\alpha_{1} = 0.2$ and $\alpha_{3} = 0.1$, respectively. The green x indicates the location of the sound source whereas the receivers are distributed uniformly across the xy-plane with the resolution of 0.3 m at the height of 1.5 m. 

\subsubsection{Results}

Fig.~\ref{fig:simulation_result_plot} shows the fitted amplitudes $\hat{A}_{k, \mathbf{x}}$ for each decay envelope at position $\mathbf{x}$ using the fade-in model. We identified the three most common decay times $T_k$ using K-means clustering.  While the amplitudes are computed for multiple octave bands, the sum over the 500, 1000, and 2000 Hz octave bands are shown here. 

Fig.~\ref{fig:simulation_result_plot}a describes the amplitudes of the shortest decay time, which is interpreted as the influence of the least reverberant room, R1. Here, the largest positive amplitudes occur in R1, as the source and the receivers are in the same room, and the lower positive amplitudes are observed in R2, as the source is now in a different room. Then, the negative amplitudes start to appear in R3, indicating the presence of fade-in reverberations. In case of Fig.~\ref{fig:simulation_result_plot}b, R3 is expected to have the largest contribution to $T_1$, as its reverberation time is in between that of R1 and R2. Thus, the amplitudes in R3 exhibit the largest positive values, while the amplitudes in R1 are negative. Lastly, Fig.~\ref{fig:simulation_result_plot}c represents the influence of the most reverberant room, R2, showing the highest positive amplitudes in R2 and less in the other two rooms. 
Note that, similar to the common-slope model, the leakage of sound between rooms is also captured with the proposed model as seen from all the transition areas near the doors.  

Additionally, we compute the root-mean-square \mbox{error~(RMSE)} between the fitted envelopes and the ground truth envelopes of the simulated RIRs. 
\begin{equation}
    RMSE_b = \sqrt{\sum_{t=1}^{L} \frac{[\bar{s}_b(\mathbf{x}, t) - s_b(\mathbf{x}, t)]^2}{L}}
\end{equation}

In Fig.~\ref{fig:simulation_error_plot}, the total RMSE across all frequency bands is depicted for both the fade-in model and the pos-only model. Since neither model incorporates the direct sound, the initial 8 ms are disregarded in the error calculation. The plots illustrate comparable errors in R1 and R2 in both models. In such cases where RIRs lack prominent fade-ins, we can see that positive amplitudes are sufficient. However, in R3, the \fadein{} model demonstrates a superior fit, because this room is where the RIRs start to exhibit pronounced fade-in characteristics. Fig.~\ref{fig:simulation_error_plot} also shows that the amount of error gets larger as the distance from the source increases. Thus, by leveraging negative amplitudes, modelling the far-away, no line-of-sight receivers becomes feasible. 



\begin{figure}[!t]
  \includegraphics[width=\columnwidth]{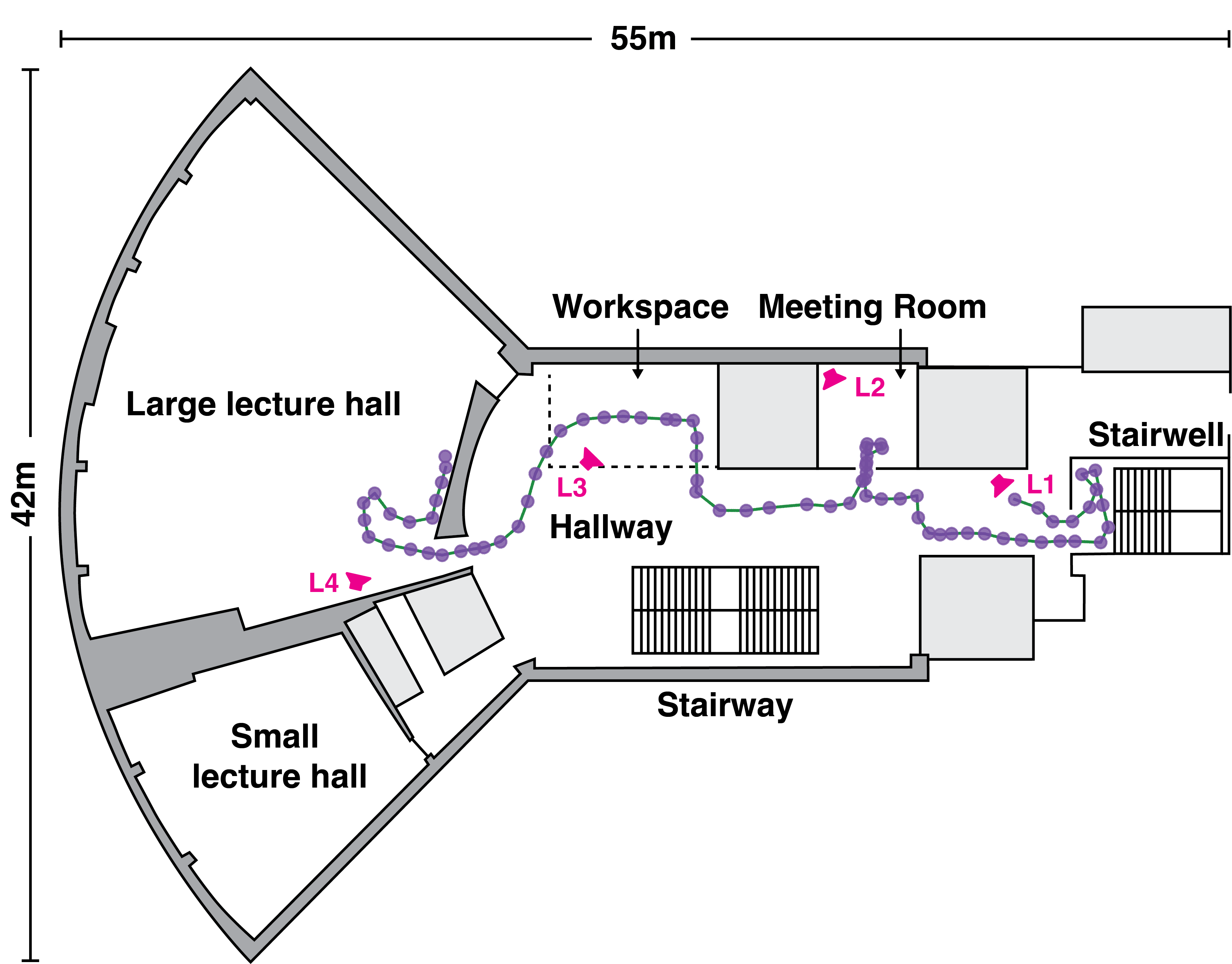}
  \caption{Map of the real-world dataset. The measurement trajectory and locations are marked with a line and circles, respectively. The locations of the loudspeakers are labeled from L1 to L4. The trajectory starts in front of L1 and a total of 77 measurements were made along the trajectory.}
  \label{fig:map}
\end{figure}

\begin{figure}
    \centering
    \subfloat[\centering Zoomed-in map of the real-world dataset]{{\includegraphics[width=0.7\columnwidth]{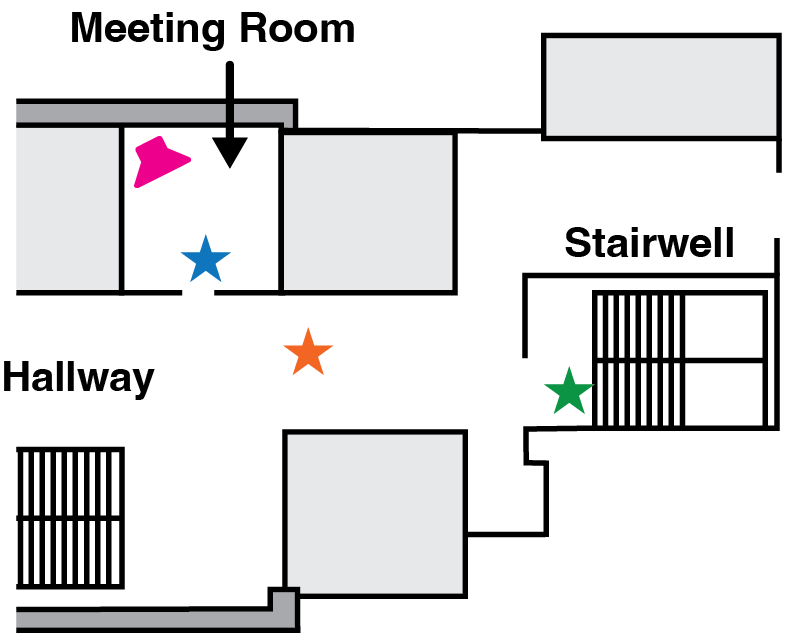} }}%
    \vfill
    \subfloat[\centering EDFs at each receiver position]{{\includegraphics[width=\columnwidth]{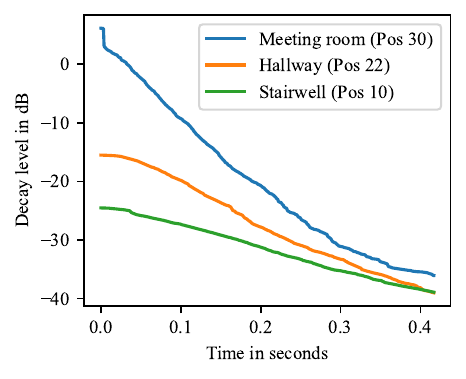} }}%
    \vfill
    \subfloat[\centering RIRs at each receiver position. Note that the y-axis of each RIR is different.]{{\includegraphics[width=\columnwidth]{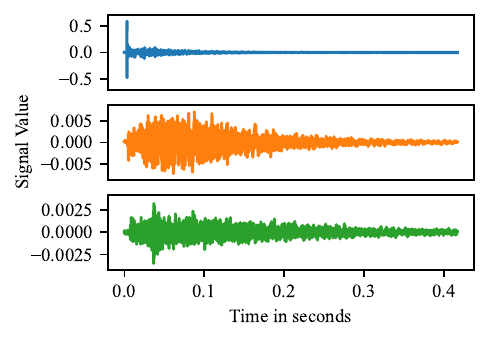} }}%

    \caption{Example of fade-in reverberations in real-world measurements. The source is in the meeting room. Three receivers in the meeting room, hallway, and stairwell are marked with stars, $\bigstar$. As with the simulation data (cf.~Fig.~\ref{fig:simulation_edf_rir}), fade-ins are observed for receivers that are not in the same room as the source.}%
    \label{fig:measurement_edf_rir}%
\end{figure}

\subsection{Real-world dataset}

\begin{figure*}[!th]
    \centering
    \subfloat[\centering L2 is active]{{\includegraphics[]{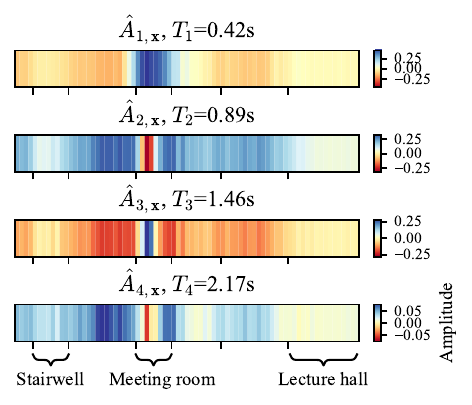} }}%
    \hspace*{-0.1em}
    \hfill
    \subfloat[\centering L4 is active]{{\includegraphics[]{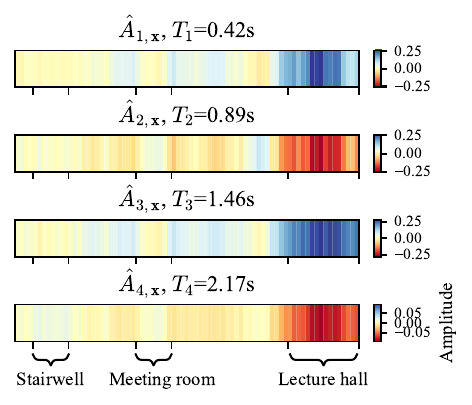} }}%
    \caption{Amplitudes along the trajectory using the \fadein{} model are shown. The left plot (a) is when the loudspeaker L2 is active and the right plot (b) is when L4 is active. Here, the amplitudes are the sum of results at the center frequency of 500 Hz, 1 kHz, and 2 kHz. As the receiver moves along the trajectory, amplitude variation is observed. Note the presence of negative amplitudes when the receiver is not in line of sight from the source.}%
    \label{fig:amplitude_transition}%
\end{figure*}

\begin{figure}[!t]
  \includegraphics[width=\columnwidth]{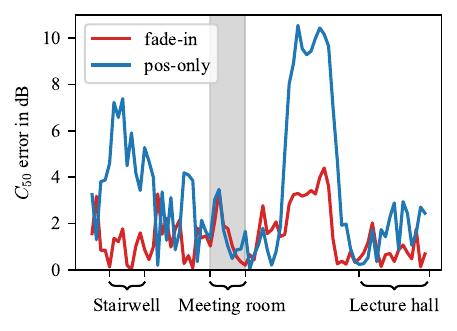}
  \caption{ $\text{C}_{50}$ error between the measured RIR vs the \fadein{} and \posonly{} RIRs. Loudspeaker 2, which is in the meeting room (shaded in grey), is active. \posonly{} model shows larger error outside the meeting room, where fade-in phenomemon occurs most prominently.}
  \label{fig:drr}
\end{figure}

\subsubsection{Dataset}
The model was also tested on a real-world dataset, which consists of RIR measurements in 3 different complex environments at Aalto University \cite{gotz2024spatially}. For this paper, we use the ``Hallway-Lecturehall'' environment, which consists of 308 SRIRs and BRIRs (77 measurement positions $\times$ 4 sources) measured along a trajectory in a coupled room environment. The map of the environment is shown in Fig.~\ref{fig:map}, where the trajectory and the measurement positions are indicated with a line and dots, respectively. 
There were 4 loudspeakers positioned in different areas. The measurement was performed continuously with looped sine sweeps from position 1 to 77, where position 1 is at one meter distance from the loudspeaker L1. 

Fig.~\ref{fig:measurement_edf_rir} shows examples of fade-in reverberation in acoustic measurements, which are similar to those observed in the simulation dataset (Fig.~\ref{fig:simulation_edf_rir}). 
With the source inside the meeting room, the receivers in the hallway and the stairwell, which are not in line of sight, exhibit the fade-in reverberation, see Figs.~\ref{fig:measurement_edf_rir}b and \ref{fig:measurement_edf_rir}c. 

\subsubsection{Results}

Fig.~\ref{fig:amplitude_transition} shows how the amplitude for each decay time varies along the trajectory for loudspeaker 2 (Fig.~7a) and loudspeaker 4 (Fig.~7b). For this dataset, four common decay times were identified per frequency band as the environment is complex with a non-trivial division of rooms. 
When the loudspeaker 2, which is in the meeting room, is active (Fig.~7a), it can be seen that $\hat{A}_{1,\mathbf{x}}$ and $\hat{A}_{3, \mathbf{x}}$ change from being negative to positive and back to negative as the receiver enters and leaves the meeting room. The opposite behavior is observed for $\hat{A}_{2,\mathbf{x}}$ and $\hat{A}_{4, \mathbf{x}}$. 
A similar trend is seen in case of the activation of loudspeaker 4 in the lecture hall (Fig.~7b). Now, when the receiver enters the lecture hall, the values of $\hat{A}_{1,\mathbf{x}}$ and $\hat{A}_{3, \mathbf{x}}$ become largely positive and $\hat{A}_{2,\mathbf{x}}$ and $\hat{A}_{4, \mathbf{x}}$ largely negative. In both figures, the areas where the amplitudes of all decay times are relatively low reflect on how the sound waves barely reach the receivers at such distant locations. Overall, the figure is a good indication that the fade-in model is capturing the effect of distance and occlusion. 

Additionally, we calculate the error in the speech clarity ($\text{C}_{50}$) \cite{iso} of RIRs synthesized using the \fadein{} and \posonly{} models. The $\text{C}_{50}$ is computed using the following equation: 
\begin{equation}
    \text{C}_{50} = 10\ \text{log}_{10} \frac{\int_{0}^{50\text{ms}}h(t)^2 dt}{\int_{50\text{ms}}^{\infty}h(t)^2 dt}.
\end{equation}
The $\text{C}_{50}$ error is defined as the difference between that of measured RIR and each of the model-synthesized RIRs. 
Fig.~\ref{fig:drr} depicts the $\text{C}_{50}$ error for both \fadein{} and \posonly{} models along the trajectory when loudspeaker 2 in the meeting room is active. Inside the meeting room, the \fadein{} and \posonly{} models exhibit similar error values, as the \fadein{} model behaves like the \posonly{} model when there is a prominent direct sound path. However, outside the room, the \posonly{} model shows larger error due to its inability to account for the absence of the direct sound path. Conversely, the \fadein{} model performs better by considering this phenomenon.


\section{Conclusion}
\label{sec:conclusion}
This work introduced the \fadein{} model, which extends the common-slope model to capture the fade-in reverberation. In order to encompass the fade-in phenomenon, which is only apparent in the initial part of RIRs, the fitting was done with the envelopes, rather than the EDFs, and negative amplitudes were allowed for the decaying exponentials. 
As a result, the evaluations on both the simulation and the real-world dataset showed that the \fadein{} model utilizes the negative amplitudes to represent RIRs at far-away and no line-of-sight locations. This was not possible with the original common-slope model. 

As a future work, we plan to render and test the model with a head-mounted device. While subjective tests have been done in the past for coupled rooms \cite{bradley2005Effects, luizard2015perceptual}, conducting one for fade-in reverberation and within a transfer-plausibility test framework \cite{wirler2020towards} has not been done to our knowledge. The effect of having visual information and a dynamic scene are also an interesting topics to explore. 


\bibliographystyle{jaes}

\bibliography{refs}

\begin{thebibliography}{25}
\newcommand{\enquote}[1]{``#1''}
\providecommand{\natexlab}[1]{#1}
\expandafter\ifx\csname urlstyle\endcsname\relax
  \providecommand{\doi}[1]{doi:\discretionary{}{}{}#1}\else
  \providecommand{\doi}{doi:\discretionary{}{}{}\begingroup \urlstyle{rm}\Url}\fi

\bibitem[{Cremer and Müller(2016)}]{CremerMuller2016PrinciplesAndApplicationsOfRoomAcoustics}
Cremer, L. and Müller, H.~A., \emph{Principles and Applications of Room Acoustics}, Peninsula Publishing, Westport, CT, USA, 2016.

\bibitem[{Kuttruff(2009)}]{Kuttruff2009RA}
Kuttruff, H., \emph{{Room Acoustics}}, CRC Press, CRC Press, 5th edition, 2009.

\bibitem[{Sü~Gül et~al.(2019)Sü~Gül, Odabaş, Xiang, and Çalışkan}]{SuGul2019DiffusionEquationModelingSoundEnergyFlowCoupled}
Sü~Gül, Z., Odabaş, E., Xiang, N., and Çalışkan, M., \enquote{{Diffusion equation modeling for sound energy flow analysis in multi domain structures},} \emph{J. Acoust. Soc. Am.}, 145(4), pp. 2703--2717, 2019, \doi{10.1121/1.5095877}.

\bibitem[{Pu et~al.(2011)Pu, Qiu, and Wang}]{pu2011different}
Pu, H., Qiu, X., and Wang, J., \enquote{Different sound decay patterns and energy feedback in coupled volumes,} \emph{J. Acoust. Soc. Am.}, 129(4), pp. 1972--1980, 2011.

\bibitem[{Svensson(1998)}]{svensson1998energy}
Svensson, U.~P., \enquote{Energy-time relations in a room with an electroacoustic system,} \emph{J. Acoust. Soc. Am.}, 104(3), pp. 1483--1490, 1998.

\bibitem[{Schr{\"o}der(2007)}]{dirk2007hybrid}
Schr{\"o}der, D., \enquote{Hybrid method for room acoustic simulation in real-time,} \emph{Proc. 19th Int. Congr. Acoust. (ICA), 2007}, 2007.

\bibitem[{Stavrakis et~al.(2008)Stavrakis, Tsingos, and Calamia}]{stavrakis2008topological}
Stavrakis, E., Tsingos, N., and Calamia, P., \enquote{Topological sound propagation with reverberation graphs,} \emph{Acta Acust. United Acust.}, 94(6), pp. 921--932, 2008.

\bibitem[{Wefers and Schr{\"o}der(2009)}]{wefers2009real}
Wefers, F. and Schr{\"o}der, D., \enquote{Real-time auralization of coupled rooms,} in \emph{Proc. EAA Auralization Symposium}, Espoo, Finland, 2009.

\bibitem[{Schr{\"o}der and Vorl{\"a}nder(2011)}]{schroder2011raven}
Schr{\"o}der, D. and Vorl{\"a}nder, M., \enquote{RAVEN: A real-time framework for the auralization of interactive virtual environments,} in \emph{Proc. Forum Acusticum}, pp. 1541--1546, Aalborg, Denmark, 2011.

\bibitem[{Piiril{\"a} et~al.(1998)Piiril{\"a}, Lokki, and V{\"a}lim{\"a}ki}]{Piirila1998}
Piiril{\"a}, E., Lokki, T., and V{\"a}lim{\"a}ki, V., \enquote{Digital signal processing techniques for non-exponentially decaying reverberation,} in \emph{Proc. 1st COST-G6 Workshop on Digital Audio Effects (DAFx)}, p. 21–24, Barcelona, Spain, 1998.

\bibitem[{Lee and Abel(2010)}]{Lee2010}
Lee, K.-S. and Abel, J.~S., \enquote{A reverberator with two-stage decay and onset time controls,} in \emph{Proc. Audio Engineering Society 129th Convention}, San Francisco, CA, 2010.

\bibitem[{Meyer-Kahlen et~al.(2020)Meyer-Kahlen, Schlecht, and Lokki}]{Meyer2020}
Meyer-Kahlen, N., Schlecht, S., and Lokki, T., \enquote{Fade-in control for feedback delay networks,} in \emph{Proc. Int. Conf. Digital Audio Effects}, pp. 227--233, 2020.

\bibitem[{Das and Abel(2021)}]{das2021grouped}
Das, O. and Abel, J.~S., \enquote{Grouped feedback delay networks for modeling of coupled spaces,} \emph{J. Audio Eng. Soc.}, 69(7/8), pp. 486--496, 2021.

\bibitem[{Kirsch et~al.(2023)Kirsch, Wendt, Van De~Par, Hu, and Ewert}]{kirsch2023computationally}
Kirsch, C., Wendt, T., Van De~Par, S., Hu, H., and Ewert, S.~D., \enquote{Computationally-efficient simulation of late reverberation for inhomogeneous boundary conditions and coupled rooms,} \emph{J. Audio Eng. Soc.}, 71(4), pp. 186--201, 2023.

\bibitem[{Xiang and Goggans(2001)}]{xiang2001evaluation}
Xiang, N. and Goggans, P.~M., \enquote{Evaluation of decay times in coupled spaces: Bayesian parameter estimation,} \emph{J. Acoust. Soc. Am.}, 110(3), pp. 1415--1424, 2001.

\bibitem[{Summers et~al.(2004)Summers, Torres, and Shimizu}]{summers2004statistical}
Summers, J.~E., Torres, R.~R., and Shimizu, Y., \enquote{Statistical-acoustics models of energy decay in systems of coupled rooms and their relation to geometrical acoustics,} \emph{J. Acoust. Soc. Am.}, 116(2), pp. 958--969, 2004.

\bibitem[{Götz et~al.(2023)Götz, Schlecht, and Pulkki}]{Goetz2023Common}
Götz, G., Schlecht, S.~J., and Pulkki, V., \enquote{Common-Slope Modeling of Late Reverberation,} \emph{IEEE/ACM Trans. Audio Speech Lang. Process.}, 31, p. 3945–3957, 2023, \doi{10.1109/TASLP.2023.3317572}.

\bibitem[{Götz et~al.(2024)Götz, Kerimovs, Schlecht, and Pulkki}]{Goetz2024DynamicAES}
Götz, G., Kerimovs, T., Schlecht, S.~J., and Pulkki, V., \enquote{Dynamic late reverberation rendering using the common-slope model,} in \emph{Proc. 6th AES Int. Conf. Audio for Games}, Tokyo, Japan, 2024.

\bibitem[{Svensson et~al.(2011)Svensson, Hassan, and Nielsen}]{svensson2011properties}
Svensson, U.~P., Hassan, E., and Nielsen, J.~L., \enquote{Properties of convolved room impulse responses,} in \emph{Proc. IEEE Workshop Appl. Signal Process. Audio Acoust.~(\mbox{WASPAA})}, pp. 205--208, IEEE, 2011.

\bibitem[{Karjalainen et~al.(2002)Karjalainen, Antsalo, M{\"a}kivirta, Peltonen, and V{\"a}lim{\"a}ki}]{karjalainen2002estimation}
Karjalainen, M., Antsalo, P., M{\"a}kivirta, A., Peltonen, T., and V{\"a}lim{\"a}ki, V., \enquote{Estimation of modal decay parameters from noisy response measurements,} \emph{J. Audio Eng. Soc.}, 50(11), pp. 867--878, 2002.

\bibitem[{G\"{o}tz et~al.(2024)G\"{o}tz, G\"{o}tz, Meyer-Kahlen, Lee, Schlecht, and Habets}]{gotz2024spatially}
G\"{o}tz, P., G\"{o}tz, G., Meyer-Kahlen, N., Lee, K.~Y., Schlecht, S.~J., and Habets, E. A.~P., \enquote{A Multi-Room Transition Dataset for Blind Estimation of Energy Decay,} \emph{18th Int. Workshop on Acoustic Signal Enhancement (IWAENC)}, 2024.

\bibitem[{ISO 3382-1:2009()}]{iso}
ISO 3382-1:2009, Acoustics — Measurement of Room Acoustic Parameters. I, 2009.

\bibitem[{Bradley and Wang(2005)}]{bradley2005Effects}
Bradley, D.~T. and Wang, L.~M., \enquote{The effects of simple coupled volume geometry on the objective and subjective results from nonexponential decay,} \emph{J. Acoust. Soc. Am.}, 118(3), p. 1480–1490, 2005, \doi{10.1121/1.1984892}.

\bibitem[{Luizard et~al.(2015)Luizard, Katz, and Guastavino}]{luizard2015perceptual}
Luizard, P., Katz, B.~F., and Guastavino, C., \enquote{Perceptual thresholds for realistic double-slope decay reverberation in large coupled spaces,} \emph{J. Acoust. Soc. Am.}, 137(1), pp. 75--84, 2015.

\bibitem[{Wirler et~al.(2020)Wirler, Meyer-Kahlen, and Schlecht}]{wirler2020towards}
Wirler, S.~A., Meyer-Kahlen, N., and Schlecht, S.~J., \enquote{Towards transfer-plausibility for evaluating mixed reality audio in complex scenes,} in \emph{Proc. AES Int. Conf. Audio for Virtual and Augmented Reality}, 2020.

\end{thebibliography}

\end{document}